\documentclass[]{aastex701}

\newcommand{\HI}{\text{H\,\textsc{I}}}
\newcommand{\fHI}{\ensuremath{f_{\rm H\,\textsc{i}}}}
\newcommand{\fHIMW}{\ensuremath{f_{\rm H\,\textsc{i},MW}}}

\begin{document}

\title{The \HI{} Mass Fractions of Milky Way Analog Samples}

\author[orcid=0000-0000-0000-0001,gname=Adam,sname='Marcello']{Adam Marcello}
\affiliation{Department of Physics and Astronomy, Haverford College, 370 Lancaster Avenue, Haverford, Pennsylvania 19041, USA}
\email[show]{marcello.tunggadjaja@gmail.com}

\author[orcid=0000-0003-0846-9578,gname=Karen, sname='Masters']{Karen L. Masters} 
\affiliation{Department of Physics and Astronomy, Haverford College, 370 Lancaster Avenue, Haverford, Pennsylvania 19041, USA}
\email{klmasters@haverford.edu}

\author[orcid=0000-0002-3746-2853,gname=David,sname='Stark']{David V. Stark} 
\affiliation{Space Telescope Science Institute, 3700 San Martin Drive, Baltimore, MD, 21218, USA}
\affiliation{William H. Miller III Department of Physics and Astronomy, Johns Hopkins University, Baltimore, MD 21218, USA}
\email{}
 
\begin{abstract}

We estimate average \HI{} mass fractions, $\fHI \equiv M_{\rm \HI}/M_\star$, for nine Milky Way analog samples (MWAs): galaxies selected on 
similarity to the MW in terms of various properties. We find most (seven) MWAs have lower $\langle$\fHI{}$\rangle$ than a recent estimate for the MW, although only the two largest have significantly lower estimates. 
We suggest that either the MW has an unusual \fHI{} for its properties, or we are overestimating its HI content, one of many challenging global measurements to make from our perspective inside our own Galaxy. 

\end{abstract}

\section{Introduction} 

Milky Way analog galaxy samples \citep[MWAs, e.g. ][L15; \citealt{Fraser-McKelvie2019}]{Licquia2015}, containing galaxies which resemble the MW, can place the properties of the MW in the context of similar galaxies in the Universe and/or provide us with valuable information about our own Galaxy that are otherwise challenging to obtain. 
The total \HI{} mass fraction, \fHI $\equiv M_{\rm \HI}/M_\star$ (\fHIMW{}) is one such property. Measurements of both total stellar and HI mass suffer from selection effects from our internal perspective \citep{Wang2020,Lian2025}. 
There are multiple ways to define MWAs. Matching with stellar mass ($M_\star$) and star formation rate (SFR) is common \citep[e.g. L15, ][]{Fraser-McKelvie2019}. \citet{Tuntipong2024} emphasize the importance of combining multiple criteria, selecting $M_{\star}$, SFR, bulge-to-total ratio ($B/T$) and effective radius, ($R_{\rm e}$). 
This research note considers the \fHI{} distributions of several MWAs and compares them with current best estimates of \fHIMW{}.

\section{Sample and Data}

Most of the MWAs we consider are from the Mapping Nearby Galaxies at Apache Point Observatory (MaNGA) DR17 (\hspace{-0.15cm}\citealt{Abdurrouf2022}); we also include some nearby galaxies from \citet[][P23]{Pilyugin2023}. For MaNGA galaxies, $M_\star$ is obtained from the GALEX-SDSS-WISE Legacy Catalog \citep[GSWLC-X2, ][]{Salim2016,Salim2018}, while \HI{} total masses and upper limits are from \HI{}-MaNGA DR4 \citep{Masters2019,Stark2021}. 
P23 provide $M_\star$ for their sample, while $M_{\rm HI}$ was sourced from various works \citep{Ball1986,Braun2009,Bush2004,Haynes2018,Koribalski2004,Koribalski2009,Mishra2012,Namumba2021,Theureau1998,Walter2008,Wang2024}. 

In total we consider nine MWAs: 
\begin{itemize}
\item two from the MaNGA MWA ancillary program - an SFR matched sample, which was designed to closely replicate L15 ($N=17$) and a $B/T$ matched sample based on \citet[][L16; $N=64$]{Licquia2016} \footnote{{\tt bitmask13} \& {\tt bitmask23}; \url{https://www.sdss4.org/dr17/manga/manga-target-selection/ancillary-targets/milky-way-analogs/}}.
    \item four MaNGA-based samples from \citet[][B20]{Boardman2020} matched in SFR ($N=6$), B/T ($N=16$), disk scale length, ($R_{\rm d}$, $N=6$), and morphology ($N=8$)
    \item $B/T$ matched MaNGA MWAs from \citet[][Z23; $N=64$]{Zhou2023} 
    \item local gas-phase metallicity 
matched MWAs (P23),  which we further limit to six galaxies with $M_\star$ consistent with $M_{\rm \star,MW}$. 
\item our own HI-MaNGA-based M$_\star$-matched subset, used to provide a reference point for \fHI{} for galaxies with stellar masses in the $M_{\rm \star,MW}$ range ($N=467$). 
\end{itemize}

\begin{figure}[bht]
\centering
\includegraphics[angle=0,scale=0.1]{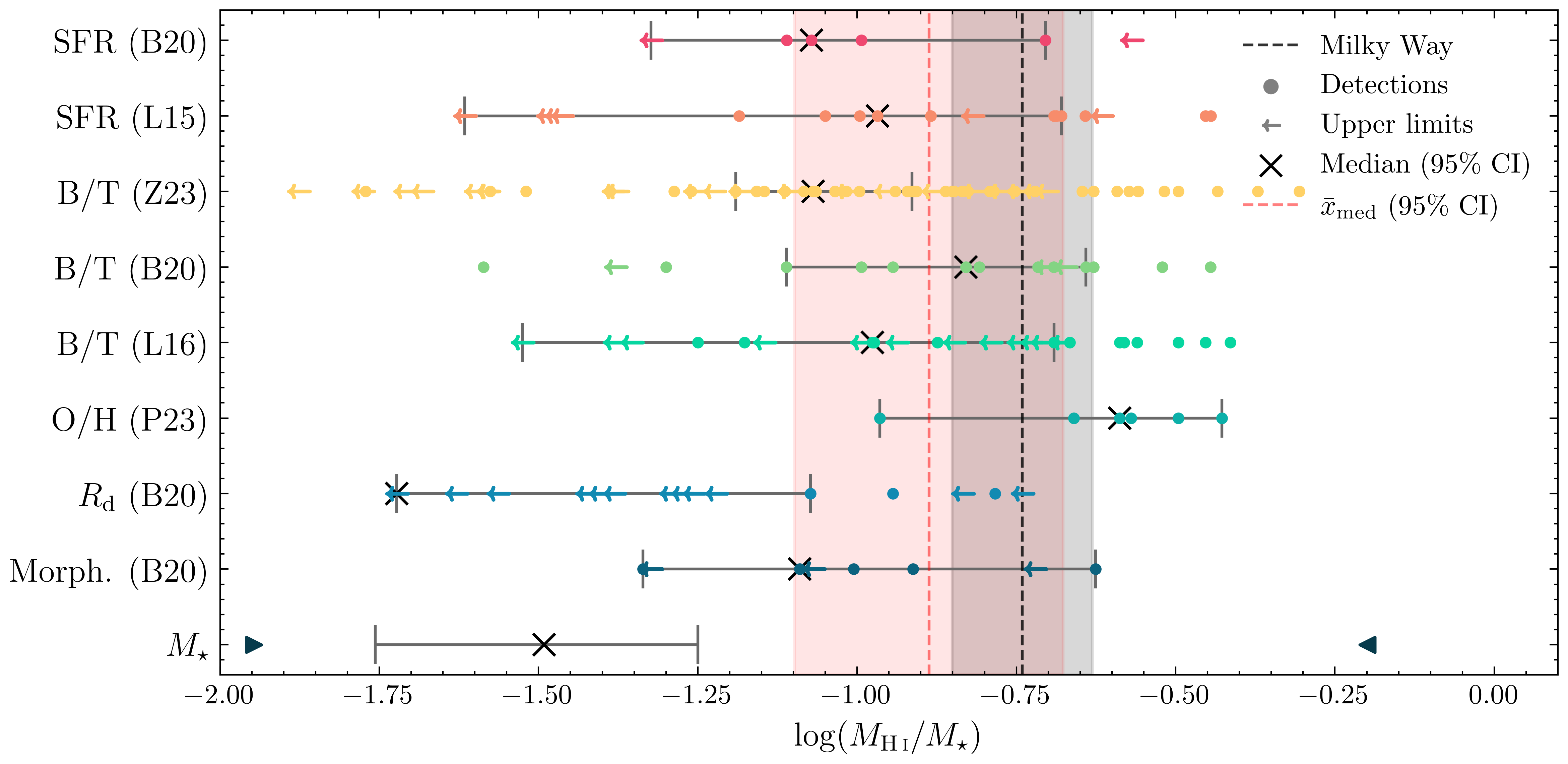}
\caption{Comparison of \fHI{} between MWAs and the MW (black dashed line and shaded CI). For all samples, we show median (crosses) and  95\% CI (horizontal lines). For most samples, individual MWA galaxy measurements are overlaid with detections (dots) and upper limits (left arrows); we show only the range (filled triangles) for the large $M_\star$-matched sample. The red dashed line show the weighted mean of all medians. 
}
\label{figure:Comparison}
\end{figure}

We compute cumulative distribution functions (CDFs) of \fHI{} for each of these MWAs including HI upper limits via the Kaplan-Meier estimator\footnote{https://github.com/dvstark/survival/tree/main}, to find the median and 95\% confidence interval (CI). These results are shown in Figure~\ref{figure:Comparison}. We adopt $\log(\fHIMW) = -0.74 \pm 0.11$, which is shown as the dashed line (grey region) based on \citet{Fielder2021} who infer $M_{\rm \star,MW} = 5.48^{+1.18}_{-0.94} \times 10^{10} \text{M}_\odot $ and \citet{Wang2020} who measure $M_{\rm \HI,MW} = (9.4\text{--}10.5) \times 10^9 \, \text{M}_\odot$. 

We find that the median \fHI{} values of most MWAs are smaller than, but agree within error of $\fHIMW{}$. 
An $R_{\rm d}$ selected MWAs (B20) has lower \fHI{}, but this may be poorly constrained due to a large fraction of upper limits. The two largest MWAs - the $B/T$-matched MWAs of Z23 and our simple mass matched MWAs have \fHI{} significantly below the MW. The inverse-variance weighted mean of all medians is $f_{\rm HI, mean}=-0.89\pm0.20$, is also just below the 95\% CI. 

\section{Discussion and Conclusion}
Most MWAs are found to have lower \fHI{} than recent estimates for \fHIMW{} \citep{Fielder2021, Wang2020}, although agree within error. The two largest MWAs have significantly lower median \fHI{} than the MW, and a weighted mean of all medians is found to be just below the 95\% CI. 
Based on this, we conclude that either the MW is unusual in its HI content, or current measurements of \fHIMW{} are overestimates. Adding to this puzzle, \citet{Lian2025} suggest current estimates of the MW stellar mass are too high, while \citet{Wang2020} note that their measurement of $M_{\rm \HI,MW}$ is an underestimate of up to 20--40\% due to a variety of factors. Both suggestions would increase \fHIMW{} relative to these MWAs.  
Even in this era of precision astrophysics, estimating global values for the MW remains challenging. 

KLM and DVS acknowledge support from the US NSF under grant AST-2510740 "Collaborative Research: Understanding the extreme diversity of atomic hydrogen depletion times in galaxies".

\newpage

\bibliography{sample701}{}

@ARTICLE{Boardman2020,
       author = {{Boardman}, N. and {Zasowski}, G. and {Newman}, J.~A. and {Andrews}, B. and {Fielder}, C. and {Bershady}, M. and {Brinkmann}, J. and {Drory}, N. and {Krishnarao}, D. and {Lane}, R.~R. and {Mackereth}, T. and {Masters}, K. and {Stringfellow}, G.~S.},
        title = "{Are the Milky Way and Andromeda unusual? A comparison with Milky Way and Andromeda analogues}",
      journal = {\mnras},
         year = 2020,
        month = nov,
       volume = {498},
       number = {4},
        pages = {4943-4954},
          doi = {10.1093/mnras/staa2731},
archivePrefix = {arXiv},
       eprint = {2009.02576},
 primaryClass = {astro-ph.GA},
       adsurl = {https://ui.adsabs.harvard.edu/abs/2020MNRAS.498.4943B}
}

@ARTICLE{Masters2019,
       author = {{Masters}, Karen L. and {Stark}, David V. and {Pace}, Zachary J. and {Phipps}, Frederika and {Rujopakarn}, Wiphu and {Samanso}, Nattida and {Harrington}, Emily and {S{\'a}nchez-Gallego}, Jos{\'e} R. and {Avila-Reese}, Vladimir and {Bershady}, Matthew and {Cherinka}, Brian and {Fielder}, Catherine E. and {Finnegan}, Daniel and {Riffel}, Rogemar A. and {Rowlands}, Kate and {Shamsi}, Shoaib and {Newnham}, Lucy and {Weijmans}, Anne-Marie and {Witherspoon}, Catherine A.},
        title = "{H I-MaNGA: H I follow-up for the MaNGA survey}",
      journal = {\mnras},
         year = 2019,
        month = sep,
       volume = {488},
       number = {3},
        pages = {3396-3405},
          doi = {10.1093/mnras/stz1889},
archivePrefix = {arXiv},
       eprint = {1901.05579},
 primaryClass = {astro-ph.GA},
       adsurl = {https://ui.adsabs.harvard.edu/abs/2019MNRAS.488.3396M}
}

@ARTICLE{Stark2021,
       author = {{Stark}, David V. and {Masters}, Karen L. and {Avila-Reese}, Vladimir and {Riffel}, Rogemar and {Riffel}, Rogerio and {Boardman}, Nicholas Fraser and {Zheng}, Zheng and {Weijmans}, Anne-Marie and {Dillon}, Sean and {Fielder}, Catherine and {Finnegan}, Daniel and {Fofie}, Patricia and {Goddy}, Julian and {Harrington}, Emily and {Pace}, Zachary and {Rujopakarn}, Wiphu and {Samanso}, Nattida and {Shamsi}, Shoaib and {Sharma}, Anubhav and {Warrick}, Elizabeth and {Witherspoon}, Catherine and {Wolthuis}, Nathan},
        title = "{H I-MaNGA: tracing the physics of the neutral and ionized ISM with the second data release}",
      journal = {\mnras},
         year = 2021,
        month = may,
       volume = {503},
       number = {1},
        pages = {1345-1366},
          doi = {10.1093/mnras/stab566},
archivePrefix = {arXiv},
       eprint = {2101.12680},
 primaryClass = {astro-ph.GA},
       adsurl = {https://ui.adsabs.harvard.edu/abs/2021MNRAS.503.1345S}
}

@ARTICLE{Salim2018,
       author = {{Salim}, Samir and {Boquien}, M{\'e}d{\'e}ric and {Lee}, Janice C.},
        title = "{Dust Attenuation Curves in the Local Universe: Demographics and New Laws for Star-forming Galaxies and High-redshift Analogs}",
      journal = {\apj},
         year = 2018,
        month = may,
       volume = {859},
       number = {1},
          eid = {11},
        pages = {11},
          doi = {10.3847/1538-4357/aabf3c},
archivePrefix = {arXiv},
       eprint = {1804.05850},
 primaryClass = {astro-ph.GA},
       adsurl = {https://ui.adsabs.harvard.edu/abs/2018ApJ...859...11S}
}

@ARTICLE{Salim2016,
       author = {{Salim}, Samir and {Lee}, Janice C. and {Janowiecki}, Steven and {da Cunha}, Elisabete and {Dickinson}, Mark and {Boquien}, M{\'e}d{\'e}ric and {Burgarella}, Denis and {Salzer}, John J. and {Charlot}, St{\'e}phane},
        title = "{GALEX-SDSS-WISE Legacy Catalog (GSWLC): Star Formation Rates, Stellar Masses, and Dust Attenuations of 700,000 Low-redshift Galaxies}",
      journal = {\apjs},
         year = 2016,
        month = nov,
       volume = {227},
       number = {1},
          eid = {2},
        pages = {2},
          doi = {10.3847/0067-0049/227/1/2},
archivePrefix = {arXiv},
       eprint = {1610.00712},
 primaryClass = {astro-ph.GA},
       adsurl = {https://ui.adsabs.harvard.edu/abs/2016ApJS..227....2S}
}

@ARTICLE{Licquia2015,
       author = {{Licquia}, Timothy C. and {Newman}, Jeffrey A. and {Brinchmann}, Jarle},
        title = "{Unveiling the Milky Way: A New Technique for Determining the Optical Color and Luminosity of Our Galaxy}",
      journal = {\apj},
         year = 2015,
        month = aug,
       volume = {809},
       number = {1},
          eid = {96},
        pages = {96},
          doi = {10.1088/0004-637X/809/1/96},
archivePrefix = {arXiv},
       eprint = {1508.04446},
 primaryClass = {astro-ph.GA},
       adsurl = {https://ui.adsabs.harvard.edu/abs/2015ApJ...809...96L}
}

@ARTICLE{Licquia2016,
       author = {{Licquia}, Timothy C. and {Newman}, Jeffrey A.},
        title = "{Sizing Up the Milky Way: A Bayesian Mixture Model Meta-analysis of Photometric Scale Length Measurements}",
      journal = {\apj},
         year = 2016,
        month = nov,
       volume = {831},
       number = {1},
          eid = {71},
        pages = {71},
          doi = {10.3847/0004-637X/831/1/71},
archivePrefix = {arXiv},
       eprint = {1607.05281},
 primaryClass = {astro-ph.GA},
       adsurl = {https://ui.adsabs.harvard.edu/abs/2016ApJ...831...71L}
}

@article{Pilyugin2023,
   title={Searching for Milky Way twins: Radial abundance distribution as a strict criterion},
   volume={676},
   ISSN={1432-0746},
   url={http://dx.doi.org/10.1051/0004-6361/202346503},
   DOI={10.1051/0004-6361/202346503},
   journal={A\&A},
   publisher={EDP Sciences},
   author={Pilyugin, L. S. and Tautvaišienė, G. and Lara-López, M. A.},
   year={2023},
   month=aug, pages={A57} }

@ARTICLE{Fraser-McKelvie2019,
       author = {{Fraser-McKelvie}, Amelia and {Merrifield}, Michael and {Arag{\'o}n-Salamanca}, Alfonso},
        title = "{From the outside looking in: what can Milky Way analogues tell us about the star formation rate of our own galaxy?}",
      journal = {\mnras},
         year = 2019,
        month = nov,
       volume = {489},
       number = {4},
        pages = {5030-5036},
          doi = {10.1093/mnras/stz2493},
archivePrefix = {arXiv},
       eprint = {1909.01654},
 primaryClass = {astro-ph.GA},
       adsurl = {https://ui.adsabs.harvard.edu/abs/2019MNRAS.489.5030F}
}

@ARTICLE{Tuntipong2024,
       author = {{Tuntipong}, Sujeeporn and {van de Sande}, Jesse and {Croom}, Scott M. and {Barsanti}, Stefania and {Bland-Hawthorn}, Joss and {Brough}, Sarah and {Bryant}, Julia J. and {Casura}, Sarah and {Fraser-McKelvie}, Amelia and {Lawrence}, Jon S. and {Ristea}, Andrei and {Sweet}, Sarah M. and {Zafar}, Tayyaba},
        title = "{The SAMI galaxy survey: on the importance of applying multiple selection criteria for finding Milky Way analogues}",
      journal = {\mnras},
         year = 2024,
        month = oct,
       volume = {533},
       number = {4},
        pages = {4334-4359},
          doi = {10.1093/mnras/stae2042},
archivePrefix = {arXiv},
       eprint = {2408.12223},
 primaryClass = {astro-ph.GA},
       adsurl = {https://ui.adsabs.harvard.edu/abs/2024MNRAS.533.4334T}
}
\bibliographystyle{aasjournalv7}

\end{document}